\documentclass[aps,prd,twocolumn,10pt,superscriptaddress,nofootinbib,nobibnotes,longbibliography]{revtex4-1}
\usepackage[final]{graphicx}
\usepackage{amsmath}
\usepackage{bbm}
\usepackage{amsfonts}
\usepackage{amssymb}
\usepackage{latexsym}
\usepackage{graphicx}
\usepackage[english]{babel}
\usepackage{multirow}
\usepackage{float}
\usepackage{url}
\usepackage{slashed}
\usepackage{xcolor}
\usepackage[utf8]{inputenc}
\usepackage{verbatim}
\usepackage{lipsum}
\usepackage{physics}
\usepackage[utf8]{inputenc}

\usepackage{diagbox}

\usepackage{mathtools}
\usepackage{amsfonts}
\usepackage{mathrsfs}
\usepackage{bbm}
\usepackage{slashed}

\usepackage{graphicx}
\usepackage{color}
\usepackage{array}
\usepackage{esint}
\usepackage{placeins}
\usepackage{booktabs}
\usepackage{makecell}
\usepackage{epstopdf}
\usepackage[caption=false]{subfig}

\usepackage{xspace}
\usepackage{siunitx}
\usepackage{hyperref}
\usepackage[nameinlink]{cleveref}
\usepackage{appendix}

\usepackage{xifthen}
\usepackage{xcolor}
\hypersetup{
	colorlinks,
	linkcolor={red!75!black},
	citecolor={blue!75!black},
	urlcolor={blue!75!black}
}
\usepackage{comment}

\newcommand{\be}{\begin{equation}}
\newcommand{\ee}{\end{equation}}
\newcommand{\ba}{\begin{array}}
\newcommand{\ea}{\end{array}}

\usepackage{ulem}
\newcommand{\beq}{\begin{eqnarray}}
\newcommand{\eeq}{\end{eqnarray}}

\begin{document}

\title{Simulating first-order phase transition during inflation}

\author{Jintao Zou}
\affiliation{Department of Physics and Chongqing Key Laboratory for Strongly Coupled Physics, Chongqing University, Chongqing 401331, China}

\author{Ligong Bian}
\email{lgbycl@cqu.edu.cn}
\affiliation{Department of Physics and Chongqing Key Laboratory for Strongly Coupled Physics, Chongqing University, Chongqing 401331, China}

\author{Shao-Jiang Wang}
\email{schwang@itp.ac.cn}
\affiliation{Institute of Theoretical Physics, Chinese Academy of Sciences (CAS), Beijing 100190, China}
\affiliation{Asia Pacific Center for Theoretical Physics (APCTP), Pohang 37673, Korea}

\begin{abstract}
Ending the inflation by vacuum decay is considered infeasible due to the graceful exit problem. Even if considering an alternative field other than the inflaton to realize a first-order phase transition (FoPT) during inflation, it is usually challenging for concrete model building, as bubble nucleations might not be fast and dense enough to successfully end the inflation. In this work, we propose a FoPT at the grand-unification-theory (GUT) scale within the Starobinsky inflation. The key construction is an exponentially evolving potential barrier dynamically controlled by the rolling inflaton, so that almost no bubble is nucleated during the early inflationary era, but with massive bubble nucleations near the end of inflation. With lattice numerical simulations, we have successfully tested this GUT-FoPT during Starobinsky inflation, and the resulting gravitational-wave energy density spectrum reproduces previous analytical estimation with a distinctive oscillation feature at high frequencies.
\end{abstract}

\maketitle

\section{Introduction} 

The inflation model has become a cornerstone of modern cosmology, providing an elegant picture for the large-scale structures on top of our flat, isotropic, and homogeneous  Universe~\cite{Brout:1977ix,Starobinsky:1980te,Kazanas:1980tx,Sato:1980yn,Guth:1980zm,Linde:1981mu,Albrecht:1982wi,Linde:1983gd}. The basic idea of inflation was first proposed explicitly in the dubbed old inflation model~\cite{Guth:1980zm} with a first-order phase transition (FoPT). However, this model suffered from the graceful-exit problem: the expansion of vacuum bubbles could not keep up with the expansion of the universe, making it impossible for inflation to end. Later, the new inflation model within the slow-roll inflation paradigm~\cite{Linde:1981mu,Linde:1983gd} allows inflation to end smoothly. Although this model fits current cosmic microwave background (CMB) observations very well~\cite{Planck:2018jri}, it often faces the challenge of ``fine-tuning'' parameters and still requires a clear physical mechanism for how the universe ``reheats'' from a supercooled phase~\cite{Bassett:2005xm,Allahverdi:2010xz,Amin:2014eta}. To seek more natural physical mechanisms, it has been widely explored that the inflaton field is coupled with other spectator fields~\cite{Linde:1990gz,Adams:1990ds,Chen:2009zp}.

One highly significant case is the coupling of the inflaton field to a field undergoing a FoPT~\cite{Ashoorioon:2015hya,Jiang:2015qor,Wang:2018caj,An:2020fff}. This coupling provides a ``graceful exit'' mechanism for inflation: the inflaton field rolls slowly during expansion, while the phase-transition field is initially trapped in a metastable vacuum. As the inflaton field evolves, it dynamically adjusts the height of the potential barrier for the phase-transition field. Eventually, the nucleation rate from the false vacuum to the true vacuum becomes large enough for the true vacuum bubbles to rapidly nucleate, expand, and collide against the background exponential expansion, releasing vacuum energy to achieve reheating and end inflation. However, the usual percolation condition with the false-vacuum fraction dropping below $1/e\approx37\%$ could be misleading during the vacuum-dominated era, as the decreasing fraction in false vacuum might not drop fast enough to compensate for the volume increase in the exponentially expanding space~\cite{Ellis:2018mja}.

To overcome this problem, we first propose in Sec.~\ref{The model set-up} a grand-unification-theory (GUT)-scale FoPT~\cite{Hu:2025xdt} during the Starobinsky inflation~\cite{Starobinsky:1980te}, where the height of the phase-transition barrier drops exponentially fast with the field excursion of the inflaton. This ensures no bubble nucleations during most of the inflationary period until the very end of the inflation. Then in Sec.~\ref{Lattice simulations}, we used 3D lattice numerical simulation techniques to simulate this two-field inflation process and obtain the corresponding gravitational wave (GW) energy density spectrum. This spectrum verifies the unique oscillation features in the high-frequency range found in Refs.~\cite{An:2020fff,An:2022cce,An:2023jxf,An:2024oui,Hu:2025xdt}, providing a distinct physical signal for future GW detection. We conclude in Sec.~\ref{sec:condis} with some discussions. Throughout Sec.~\ref{Lattice simulations}, we adopt units where $c=8\pi G=1$.

\section{The model set-up}
\label{The model set-up}
In this section, we first propose a phenomenological model for a GUT-scale FoPT during the Starobinsky inflation, followed by a supergravity (SUGRA)-inspired effective-field-theory (EFT) model-building, and then we detailize successful configurations of FoPT completion for later input in the numerical simulations. 

\subsection{A GUT FoPT during Starobinsky inflation}

\begin{figure}
    \centering
    \includegraphics[width=0.48\textwidth]{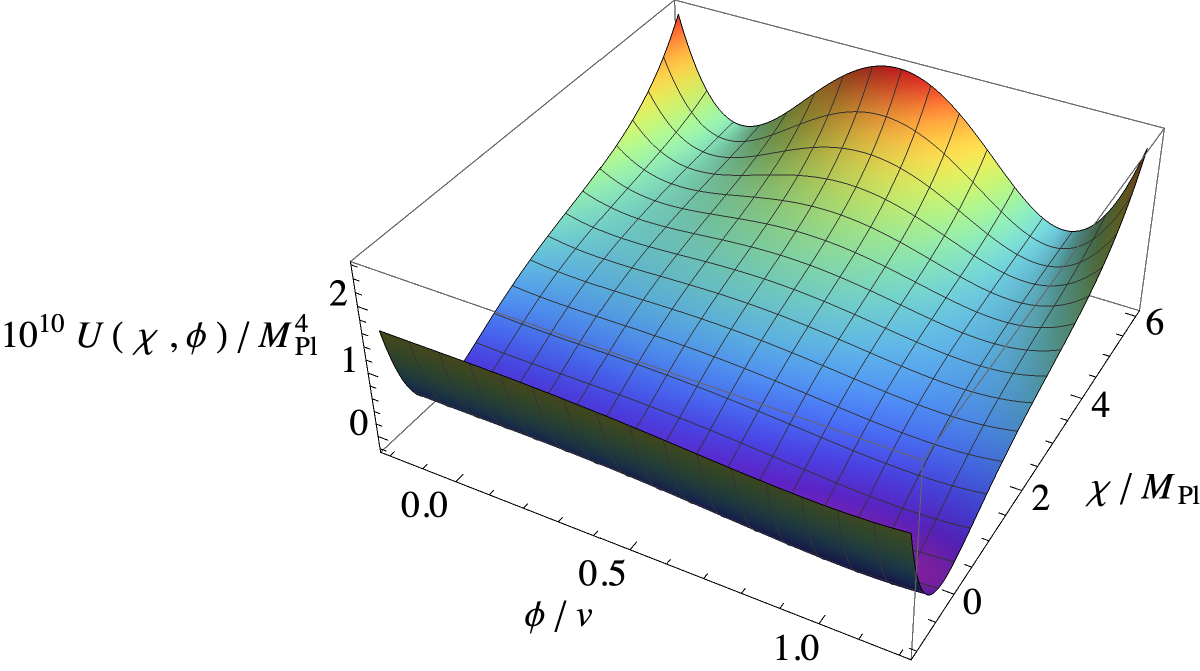}\\
    \caption{A 3D visualization of the total potential.}
    \label{fig:potential}
\end{figure}

We first propose a phenomenological model as follows,
\begin{equation}
    S=\int \mathrm{d}^4 x \sqrt{-g}\left[\frac{M_{\mathrm{Pl}}^2}{2} R-\frac{1}{2}(\nabla \chi)^2-\frac{1}{2}(\nabla \phi)^2-U(\chi, \phi)\right]
\end{equation}
where $M_{\mathrm{Pl}}$ is the reduced Planck mass, the total potential $U(\chi, \phi)=W(\chi)+V(\phi)$ consists of a plateau potential $W(\chi)$ as in the Starobinksy inflation~\cite{Starobinsky:1980te,Kehagias:2013mya,Whitt:1984pd,Maeda:1987xf,Gorbunov:2010bn}, and a FoPT potential $V(\phi)$ with the height of potential barrier controlled by the parameter $a$,
\begin{equation}
W(\chi)=\alpha M_{\mathrm{Pl}}^4\left(1-e^{-\beta \chi / M_{\mathrm{Pl}}}\right)^2, \quad \beta=\sqrt{2 / 3},
\end{equation}
\begin{equation}\label{eq:V}
V(\phi)=\gamma M_{\mathrm{Pl}}^4\left[a\left(\frac{\phi}{v}\right)^2-(2 a+4)\left(\frac{\phi}{v}\right)^3+(a+3)\left(\frac{\phi}{v}\right)^4\right].
\end{equation}
Here, the FoPT potential is defined at a GUT scale $v\equiv\gamma^{1/4}M_\mathrm{Pl}$, which should be slightly below the inflationary scale $W^{1/4}_\mathrm{inf}\simeq\alpha^{1/4}M_\mathrm{Pl}$ so as not to jeopardize the early-stage inflationary era. The polynomial coefficients $(a,2a+4,a+3)$ are chosen in such a way that the false and true vacua are located exactly at $V(\phi/v=0,1)=0,-v^4$, respectively, and the barrier disappears exactly at $a=0$. The rolling inflation dynamically controls the barrier height via 
\begin{align}
a=e^{\beta\chi/M_\mathrm{Pl}}-1,
\end{align}
which is shown in Fig.~\ref{fig:potential} with an illustrative parameter choice $\alpha=10^{-10}=5\gamma$.

We then give rise to an SUGRA-inspired EFT model for a possible ultraviolet (UV) completion of the above two-field potential~\cite{Ellis:2013xoa,Kallosh:2013lkr,Cecotti:1987sa,Ferrara:2013rsa,Buchmuller:2013zfa}.
The inflaton sector of the Starobinsky inflation follows the same SUGRA construction as the Cecotti  model~\cite{Cecotti:1987sa} and its no-scale SUGRA realization~\cite{Ellis:2013nxa}. In the $\mathcal{N}=1$, $D=4$ SUGRA with multiple chiral superfields (complex scalar $\phi^i$ and its complex-conjugate $\bar{\phi}^{\bar{i}}$), the effective potential $\mathcal{V}$ is constructed from a K\"{a}hler potential $\mathcal{K}$ and a superpotential $\mathcal{W}$ as
\begin{align}
\mathcal{V}=e^{\mathcal{K}}\left(\mathcal{K}^{i\bar{j}}\mathcal{D}_i\mathcal{W}\mathcal{D}_{\bar{j}}\bar{\mathcal{W}}-3|\mathcal{W}|^2\right), 
\end{align}
where 
\begin{align}
\mathcal{D}_i\mathcal{W}=\frac{\partial\mathcal{W}}{\partial\phi^i}+\frac{\partial \mathcal{K}}{\partial\phi^i}\mathcal{W},\quad 
\mathcal{K}_{i\bar{j}}=\frac{\partial^2\mathcal{K}}{\partial\phi^i\partial\phi^{\bar{j}}}.
\end{align}
The chiral superfields of the Starobinsky inflation consist of a Starobinsky modulus $\mathcal{T}$ and an inflation stabilizer $\mathcal{S}$ with the K\"{a}hler/superpotential potentials of forms
\begin{align}
\mathcal{K}_\mathrm{inf}=-3\ln\left(\mathcal{T}+\bar{\mathcal{T}}-|\mathcal{S}|^2\right),\quad
\mathcal{W}_\mathrm{inf}=M\mathcal{S}(\mathcal{T}-\frac12),
\end{align}
where the dimensionless $M$ will be determined later. When $\mathcal{S}$ is stablized at $\mathcal{S}=0$, it recovers the Starobinsky plateau potential,
\begin{align}
\mathcal{V}_\mathrm{inf}=\frac{M^2}{3}\frac{|\mathcal{T}-1/2|^2}{(\mathcal{T}+\bar{\mathcal{T}})^2}=\frac{M^2}{12}\left(1-e^{-\sqrt{\frac23}\chi}\right)^2,
\end{align}
after fixing $M=2\sqrt{3\alpha}$ and defining $\mathcal{T}=\frac12e^{\sqrt{\frac23}\chi/M_\mathrm{Pl}}+ia$ in terms of the Starobinsky scalaron $\chi$ and axion $a$ (stablized at $a=0$). 

The embedding of a FoPT potential~\eqref{eq:V} into the no-scale SUGRA construction of Starobinsky inflation with another chiral field $\Phi\equiv\phi/\sqrt{2}$ is highly non-trivial, as our coefficient $a=e^{\beta\chi}-1=\mathcal{T}+\bar{\mathcal{T}}-1$ depends explicitly on $\bar{\mathcal{T}}$, whose presence in the superpotential would not be holomorphic anymore. This is why most of the naive SUGRA constructions would fail in the end. The other difficulty is the cubic term in $V$ that would require non-trivial cancellations between two terms in $\mathcal{V}$. From an EFT point of view, $a$ changes with the background $\chi$ and can therefore be treated as a spurion field. Hence, we propose the K\"{a}hler potential and superpotential as
\begin{align}
\mathcal{K}&=-3\ln(\mathcal{T}+\bar{\mathcal{T}}-|\mathcal{S}|^2-|\Phi|^2),\\
\mathcal{W}&=\mathcal{W}_\mathrm{inf}+\gamma M_\mathrm{Pl}^4[c_2(a)\Phi^2+c_3(a)\Phi^3+c_4(a)\Phi^4]),
\end{align}
where the coefficients are of the following forms,
\begin{align}
c_2(a)&=\frac{\sqrt{3a}}{2}(a+1),\\
c_3(a)&=-\frac{1}{\sqrt{3a}}(a+2)(a+1),\\
c_4(a)&=-\frac{\sqrt{3}}{128a^{3/2}}(45a^2+82a+65),
\end{align}
to match the coefficients in $V$.
This might appear fine-tuning at first sight, partially because our coefficients in $V(\phi)$ are organized in such a way that the false and true vacua are located exactly at $V(\phi/v=0,1)/v^4=0,-1$, respectively, and the barrier disappears exactly at $a=0$. A more arbitrary choice of coefficients in $V(\phi)$, but to maintain the overall barrier shape, would greatly relieve the fine-tuning in a concrete SUGRA construction.

The equations of motion (EoM) of this two-field evolution read
\begin{equation}
\begin{aligned}
\frac{1}{2} \dot{\chi}^2+\frac{1}{2} \dot{\phi}^2+\frac{(\vec{\nabla} \chi)^2+(\vec{\nabla} \phi)^2}{2 a^2}+U &= 3 M_{\mathrm{Pl}}^2 H^2, \\
-\frac{1}{2} \dot{\chi}^2-\frac{1}{2} \dot{\phi}^2-\frac{(\vec{\nabla} \chi)^2+(\vec{\nabla} \phi)^2}{6 a^2} &= M_{\mathrm{Pl}}^2 \dot{H}, \\
\ddot{\chi}-\frac{\vec{\nabla}^2 \chi}{a^2}+3 H \dot{\chi}+\frac{\partial U}{\partial \chi}&=0 \\
\ddot{\phi}-\frac{\vec{\nabla}^2 \phi}{a^2}+3 H \dot{\phi}+\frac{\partial U}{\partial \phi}&=0,
\label{Eq:EoM}
\end{aligned}
\end{equation}
which can be made dimensionless as follows,
\begin{align}
    \frac{\tilde{H}^2}{2}\left(\tilde{\chi}^{\prime 2}+\tilde{\phi}^{\prime 2}\right)+\frac{e^{2 N}}{2}\left[(\tilde{\nabla} \tilde{\chi})^2+(\tilde{\nabla} \tilde{\phi})^2\right]+\tilde{U}&= 3 \tilde{H}^2,\\
    \frac{\tilde{H}^2}{2}\left(\tilde{\chi}^{\prime 2}+\tilde{\phi}^{\prime 2}\right)+\frac{e^{2 N}}{6}\left[(\tilde{\nabla} \tilde{\chi})^2+(\tilde{\nabla} \tilde{\phi})^2\right] &= \tilde{H} \tilde{H}^{\prime},\\
    \tilde{\chi}^{\prime \prime}+\frac{\tilde{H}^{\prime}}{\tilde{H}} \tilde{\chi}^{\prime}-\frac{e^{2 N}}{\tilde{H}^2} \tilde{\nabla}^2 \tilde{\chi}-3 \tilde{\chi}^{\prime}+\frac{1}{\tilde{H}^2} \frac{\partial \tilde{U}}{\partial \tilde{\chi}}&=0,\\
    \tilde{\phi}^{\prime \prime}+\frac{\tilde{H}^{\prime}}{\tilde{H}} \tilde{\phi}^{\prime}-\frac{e^{2 N}}{\tilde{H}^2} \tilde{\nabla}^2 \tilde{\phi}-3 \tilde{\phi}^{\prime}+\frac{1}{\tilde{H}^2} \frac{\partial \tilde{U}}{\partial \tilde{\phi}}&=0,
\end{align}
after abbreviating $\tilde{H} \equiv H/M_{\mathrm{Pl}}$, $\tilde{\chi} \equiv \chi/M_{\mathrm{Pl}}$, $\tilde{\phi} \equiv \phi/M_{\mathrm{Pl}}$, $\tilde{U} \equiv U/M_{\mathrm{Pl}}^4 $, and $\tilde{H}^{\prime} \equiv \mathrm{d} \tilde{H}/\mathrm{~d} N$, $\tilde{\chi}^{\prime} \equiv \mathrm{d} \tilde{\chi}/\mathrm{~d} N$, $\tilde{\phi}^{\prime} \equiv \mathrm{d} \tilde{\phi}/\mathrm{~d} N$, $\tilde{x} \equiv x M_{\mathrm{Pl}}$, $\tilde{y} \equiv y M_{\mathrm{Pl}}$, $\tilde{z} \equiv z M_{\mathrm{Pl}}$.

\subsection{Model configurations for lattice simulations}

We now turn to set up the initial configurations for later lattice simulation. Before that, we have explicitly checked whether quantum fluctuations $\delta\phi\simeq H_\mathrm{inf}/2\pi\simeq\sqrt{W(\chi)/3M_\mathrm{Pl}^2}$ during inflation could overcome the potential barrier $V(\phi_\mathrm{max})^{1/4}$ before bubbles ever nucleate, where $\phi_\mathrm{max}$ is the field value at barrier peak. It is easy to verify $H_\mathrm{inf}/2\pi\ll V(\phi_\mathrm{max})^\frac14$ during inflation with $\mathcal{O}(0.1)\lesssim\chi/M_\mathrm{Pl}<\mathcal{O}(1)$, that is,
\begin{align}
\sqrt{\frac{\alpha}{3}}\frac{(1-e^{-\beta\chi})}{2\pi}\ll\frac{\gamma^\frac14}2\left(\frac{(e^{\beta\chi}+3)(e^{\beta\chi}-1)^3}{(e^{\beta\chi}+2)^3}\right)^\frac14.
\end{align}

The first stage evolution is the usual slow-roll inflation. The potential slow-roll parameter,
\begin{equation}
    \epsilon_V = \frac{M_\mathrm{Pl}^2}{2} \left( \frac{W'(\chi)}{W(\chi)} \right)^2 = \frac{2 \beta^2}{(e^{\beta \chi / M_\mathrm{Pl}} - 1)^2},
\end{equation}
when saturating $\epsilon_V(\chi_0)=1$, fixes the field value $\chi_0=(M_\mathrm{Pl}/\beta)\ln(1+\sqrt{2}\beta)=0.94M_\mathrm{Pl}$ at the end of the slow-roll inflation. Here, we count the e-folding number $N$ backward in time via $\mathrm{d}N=-H\mathrm{d}t=-(H/\dot{\chi})\mathrm{d}\chi=-\mathrm{d}\chi/M_\mathrm{Pl}/\sqrt{2\epsilon_H}\approx-\mathrm{d}\chi/M_\mathrm{Pl}/\sqrt{\epsilon_V}$. The e-folding number can be explictily integrated as
\begin{equation}
    N(\chi)\approx \int_{\chi_{\mathrm{end}}}^{\chi} \frac{W(\chi')}{W'(\chi')} \frac{\mathrm{d}\chi'}{M_{\mathrm{Pl}}^2} = \frac{e^{\beta \chi / M_\mathrm{Pl}} - \beta \chi / M_\mathrm{Pl} - \mathcal{C}}{2\beta^2},
\end{equation}
where the integration constant $\mathcal{C} = 1 + \sqrt{2}\beta - \ln(1 + \sqrt{2}\beta)$ ensures that $N(\chi_0) = 0$. Further introducing the Lambert $W$ function (specifically the $W_{-1}$ branch), the field value can be inverted analytically as
\begin{align}
    \frac{\chi_N}{M_\mathrm{Pl}} = & -\frac{1}{\beta} \Big[ 1 + \beta(\sqrt{2} + 2N\beta) - \ln(1 + \sqrt{2}\beta) \nonumber\\
    & + W_{-1} \left( -e^{-1 - \beta(\sqrt{2} + 2N\beta)} (1 + \sqrt{2}\beta) \right) \Big].
\label{Eq:chi}
\end{align}
With this solution, we can further compute the Hubble parameter and scalar spectrum amplitude as
\begin{align}
\frac{H^2(N)}{M_\mathrm{Pl}^2}&=\frac{W(\chi_N)/M_\mathrm{Pl}^4}{3\left(1-\frac16\frac{\chi_N^2}{M_\mathrm{Pl}^2}\right)},\\
A_s(N)&=\frac{1}{4\pi^2}\frac{H^2(N)}{\chi_N^2}.
\end{align}
For $N=60$, it admits $\chi_{60}=5.45M_\mathrm{Pl}$, and $\alpha=0.94\times10^{-10}$ for $A_s(N=60)$ to match the observed $A_s=2.1\times10^{-9}$~\cite{Planck:2018jri}, otherwise we can choose $N=58.21$ to reach the same matching for $\alpha=1.0\times10^{-10}$.

\begin{figure}
    \centering
    \includegraphics[width=0.48\textwidth]{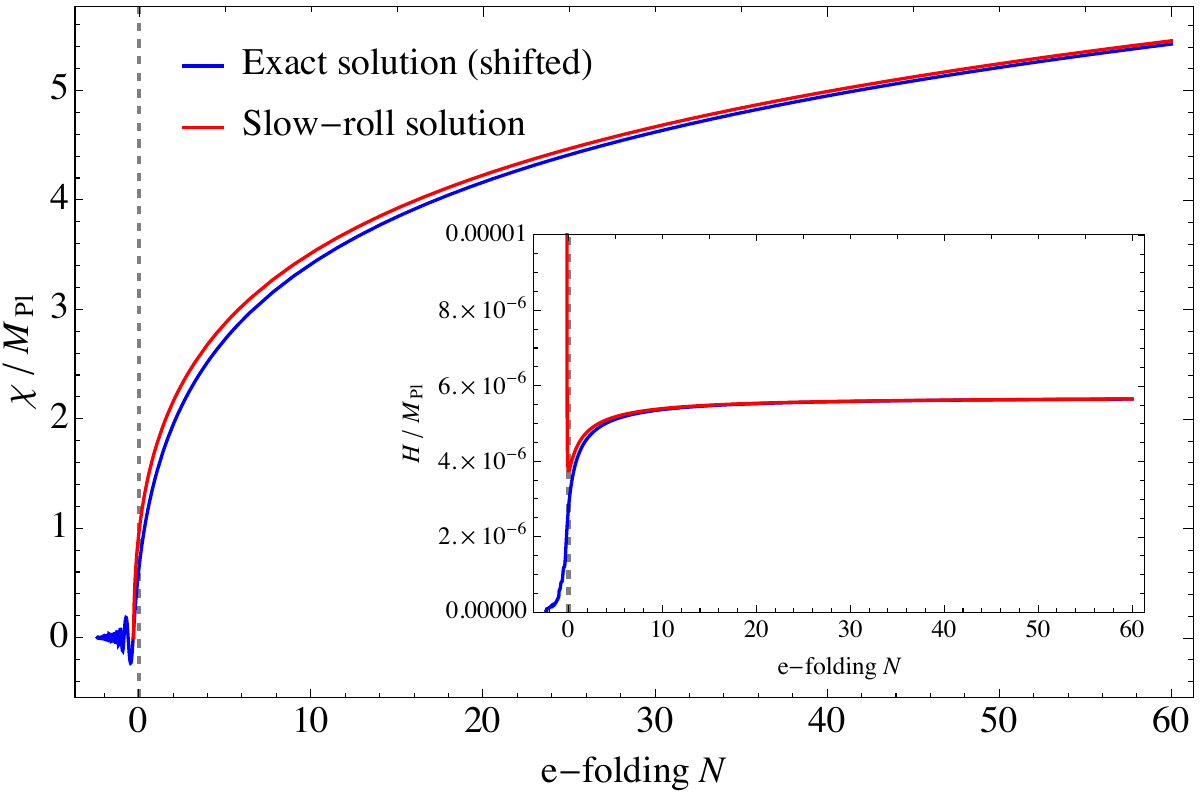}\\
    \caption{Comparisons of inflaton field value and Hubble parameter evolutions between the approximated slow-roll solution and the exact solution shifted to match $\epsilon_H(N=0)=1$.}
    \label{fig:inflation}
\end{figure}

However, the above analytical estimation is not exact as we have approximated $\epsilon_H=-\dot{H}/H^2$ with $\epsilon_V$. The exact solution can be numerically solved from combining $\chi$-EoM with two FLRW equations for $H^2$ and $\dot{H}$ as
\begin{align}
\tilde{\chi}''+\frac12\tilde{\chi}'^3-3\tilde{\chi}'+3(1-\frac16\tilde{\chi}'^2)\frac{\tilde{W}_\chi}{\tilde{W}}=0.
\end{align}
With this exact numerical solution as shown in Fig.~\ref{fig:inflation}, the saturation of $\epsilon_H(\chi_\mathrm{end}\equiv\chi(N_\mathrm{end}))=1$ gives rise to $N_\mathrm{end}=-1.52$ and $\chi_\mathrm{end}=0.61M_\mathrm{Pl}$. To match our convention that the original slow-roll inflation ends exactly at $N=0$, we can shift the e-folding number by $N_\mathrm{end}$, in which case $A_s(N=60-N_\mathrm{end})=2.1\times10^{-9}$ gives rise to $\alpha=0.98\times10^{-10}$ that leads to $H_\mathrm{CMB}(N=60-N_\mathrm{end})=5.65\times10^{-6}M_\mathrm{Pl}$ and $\chi_\mathrm{CMB}(N=60-N_\mathrm{end})=5.42M_\mathrm{Pl}$.

The second stage evolution is for some patches to tunnel through the potential barrier with sufficient nucleations of true-vacuum bubbles to quickly end the inflation, but before that, the potential barrier was too high to nucleate any/enough bubbles. To make concrete of this picture, note that, for bubbles nucleated at $t_i$ and collided at $t$ during inflation with a near-constant Hubble scale $H$ and an e-folding number $N-N_i\approx-H(t-t_i)$, the comoving distance that bubble wall sweeps over reads
\begin{align}
r(t,t_i)=\int_{t_i}^t\frac{\mathrm{d}\tilde{t}}{a(\tilde{t})}
\approx\frac{1-e^{N-N_i}}{a(t_i)H},
\end{align}
where the scale factor evolves as $a(t)\approx a(t_i)e^{H(t-t_i)}$. For bubbles with averaged initial physical separation $d(t_i)$,
the necessary condition for bubbles to collide at time $t$ is
\begin{align}
\frac{d(t_i)}{a(t_i)}\equiv\frac{1}{a(t_i)n_b(t_i)^{1/3}}\leq r(t,t_i)\equiv\frac{1-e^{N-N_i}}{a(t_i)H},
\end{align}
so that the percolation time scale $t-t_i=d(t_i)$ should be smaller than the Hubble time scale,
\begin{align}
H^{-1}\geq\frac{d(t_i)}{1-e^{N-N_i}}>d(t_i).
\end{align}
that is, if we cannot end the inflation with bubble collisions within one e-folding number, we can never end it forever ever since then. This observation further yields the physical number density $n_b(t)$ to be
\begin{align}
\frac{n_b(t_i)}{H^3}\geq\frac{1}{(1-e^{N-N_i})^3}=\frac{1}{(1-e^{-1})^3}\approx4,
\label{Eq:n_b}
\end{align}
that is, the number of bubbles in one comoving Hubble volume must reach 4 in order to collide within one e-folding number.

\begin{figure}
    \centering
    \includegraphics[width=0.5\textwidth]{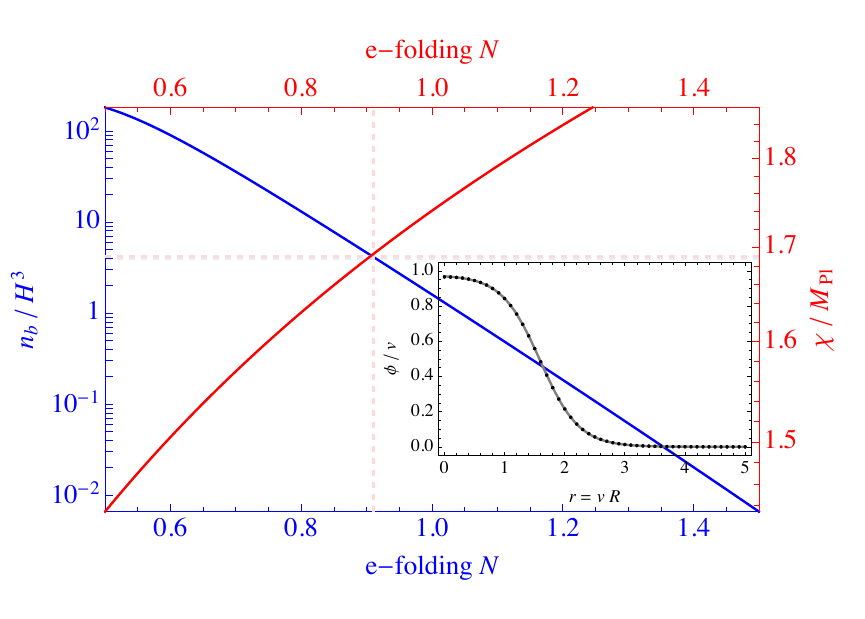}
    \caption{The number of bubbles per unit comoving Hubble volume (blue) and the corresponding inflaton $\chi$-field value (red) as functions of the e-folding number $N$, where the intersection point presents the necessary condition for bubble collisions within one e-folding number. The inset presents the fitted bubble profile nucleated at the time corresponding to the intersection point.}
    \label{fig:NumberDensity}
\end{figure}

To estimate the time evolution of the number density of bubbles, recall that it is defined via $\frac{\mathrm{d}}{\mathrm{d}t}[n_b(t)a(t)^3]=\Gamma(t)F(t)a(t)^3$. Here, the vacuum decay rate $\Gamma(t)=R_0^{-4}e^{-S_4}$ per unit time and per unit volume depends on the initial bubble radius $R_0$ and its Euclidean action $S_4[\phi_B(r)]$ evaluated at its $O(4)$ bounce solution $\phi_B(r)$ from public code \texttt{AnyBubble}~\cite{Masoumi:2016wot}, where the initial bubble radius $R_0$ is defined as $\phi_B(R_0)=\phi_B(r=0)/2$. The false-vacuum fraction $F(t)=e^{-f(t)}$ is calculated from $f(t)=\frac{4\pi}{3}\int_0^t\mathrm{d}t'\,\Gamma(t')a(t')^3r(t,t')^3$~\cite{Guth:1982pn,Turner:1992tz}. All of the above quantities are functions of $\chi(N)$ when vacuum decay in $\phi$ direction occurs, and hence we can calculate
\begin{align}
\frac{n_b(N)}{H(N)^3}&=\frac{1}{[H(N)R_0(N)]^4}\int_N^\infty\mathrm{d}N'e^{-S_4(N')-f(N')+3(N-N')},\\
f(N)&=\frac{4\pi}{3[H(N)R_0(N)]^4}\int_N^\infty\mathrm{d}N'e^{-S_4(N')}(1-e^{N-N'})^3.
\end{align}

In Fig.~\ref{fig:NumberDensity}, we present the comoving number density of bubbles as a function of $\chi(N)$, where the necessary condition for bubble collisions during inflation is fulfilled around $\chi_*=1.69 M_\mathrm{Pl}$ at an e-folding number $N_*=0.91$, around which the necessary percolation condition $|f'(N_*)|\approx339\gg3$~\cite{Ellis:2018mja} has been chekced. The bubble profile at this point can be numerically fitted as
\begin{align}
\phi_B(r)/v=0.488\left(1-\mathrm{tanh}\left(\frac{v r-1.594}{0.643}\right)\right).
\end{align}
Here, we take $\alpha = 10^{-10}$, $\beta = \sqrt{2/3}$, $\gamma = 0.2 \times 10^{-10}$, and hence $v=\gamma^{1/4}M_\mathrm{Pl}=5.15\times10^{15}\mathrm{GeV}$ is at GUT scale. Note that the traditional percolation condition $F(t_\mathrm{per})=e^{-1}$ or $70\%$~\cite{percolation1971} is insufficient for successful bubble collisions during inflation since false-vacuum fraction could still $F(t)\to0$ even if bubbles never collide~\cite{Ellis:2018mja}.

\section{Lattice simulations}
\label{Lattice simulations}
In this section, we first introduce the evolution equations and parameter settings used for the simulation, followed by a detailed presentation of the corresponding numerical results.

\subsection{Simulation setup}
\label{Simulation setup}
We discretize the evolution equations on a three-dimensional uniform Cartesian grid with periodic boundary conditions. The resulting coupled system of equations is integrated forward in time using the fourth-order Runge-Kutta method provided in the {\it pystella}~\cite{weiner2021stencil,Adshead:2019igv,Adshead:2019lbr,Weiner:2021zj}\footnote{https://github.com/zachjweiner/pystella.}.
The classical equations of motion governing the FLRW background evolution are given by
\begin{equation}
\begin{aligned}
    \mathcal{H}(\tau)&=a^{\prime}(\tau)/a(\tau)=\sqrt{\frac{a(\tau)^2}{3M^2_\mathrm{Pl}}\rho_\mathrm{tot}},
\end{aligned}
\end{equation}
where $\tau$ denotes conformal time. The two fields evolve according to the Klein-Gordon equation Eq.~(\ref{Eq:EoM}) in an expanding universe. The average energy density and pressure are given by
\begin{equation}
\begin{aligned}
    \rho_\mathrm{tot} &= \rho_\mathrm{K} + \rho_\mathrm{G} + \rho_\mathrm{V} \\
    p &= \rho_\mathrm{K} - \frac{1}{3}\rho_\mathrm{G} - \rho_\mathrm{V},
\end{aligned}
\end{equation}
where $\rho_\mathrm{K}=\phi^{\prime 2}/2$, $\rho_\mathrm{G}=(\partial_i \phi)^2/2a^2$, and $\rho_\mathrm{V}$ represent the spatially averaged kinetic energy density, gradient energy density, and potential energy density.

As bubble collisions become more frequent, the initially homogeneous spatial configuration is progressively disrupted, leading to an inhomogeneous and anisotropic energy distribution. The bubble walls, which store a significant amount of energy, release it during collisions, thereby generating pronounced anisotropic perturbations. These anisotropies break the spherical symmetry of the system and induce strong oscillations in the spacetime curvature, which propagate through the universe as gravitational waves (GWs). 
To investigate GW generation, we calculate the tensor perturbation ${h}_{\mu\nu}$ at all times, 
\begin{equation}
    h''_{\mu\nu} - \nabla^2 h_{\mu\nu} = 16\pi G T_{\mu\nu},
\end{equation}
where the $T_{\mu\nu}$ is the energy-momentum tensor of the scalar field $\phi$ and only the transverse-traceless part of the energy-momentum tensor acts as the source. The effective energy-momentum tensor and the corresponding energy density of the GWs are given by~\cite{Hindmarsh:2015qta,Garcia-Bellido:2007fiu}:
\begin{equation}
\begin{aligned}
T_{\mu\nu}^{\text{GW}} &= \frac{1}{32 \pi G} \langle \partial_\mu h_{ij} \partial_\nu h^{ij} \rangle, \\
\rho_{\text{GW}} &= \frac{1}{32\pi G} \langle h'^2_{ij} \rangle.
\end{aligned}
\end{equation}
The GW power spectrum is then expressed as
\begin{equation}
    \Omega_{\text{GW}}(k) = \frac{1}{24\pi^2\mathcal{H}^2 V} \sum_{i,j} \int \mathrm{d}\Omega |\mathbf{k}|^3 | h'_{ij}(k,\tau) |^2,
\end{equation}
where $V$ is the simulation volume.

\begin{figure}[!htp]
\centering
    \includegraphics[width=0.23\textwidth]{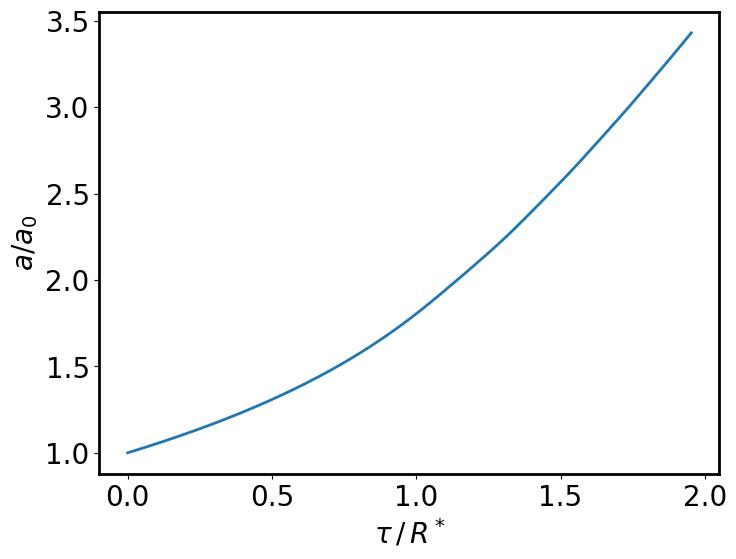}
    \includegraphics[width=0.23\textwidth]{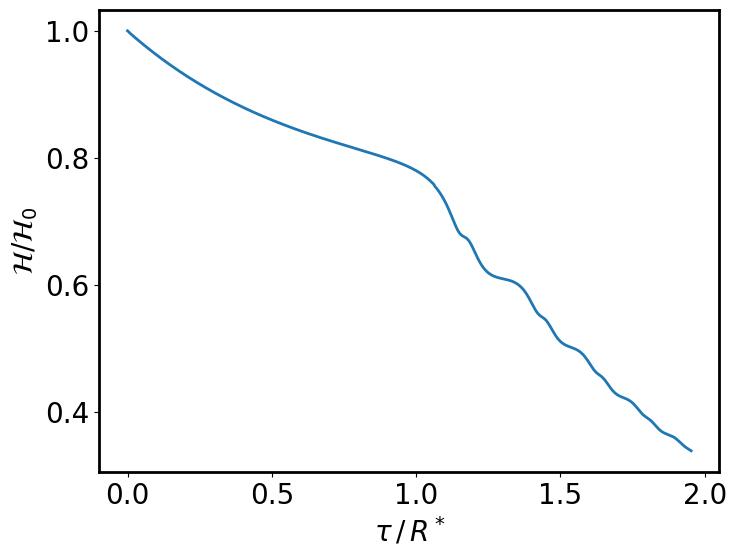}
    \includegraphics[width=0.23\textwidth]{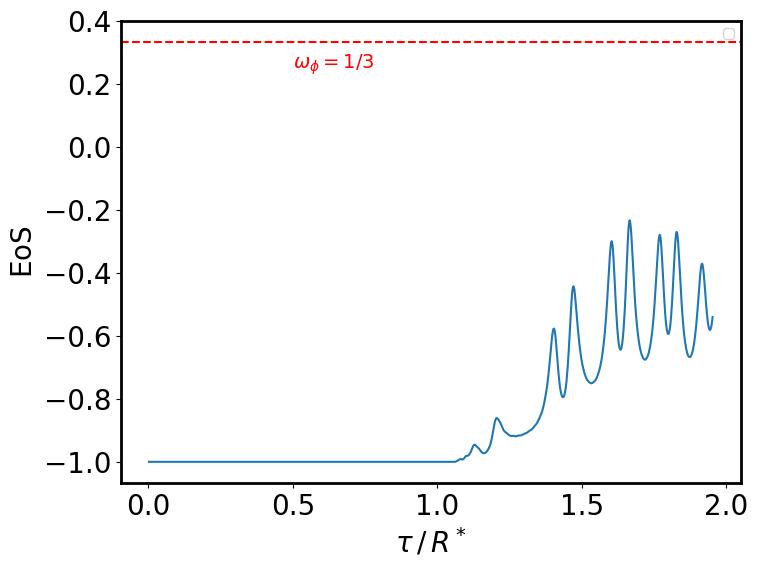}
    \includegraphics[width=0.24\textwidth]{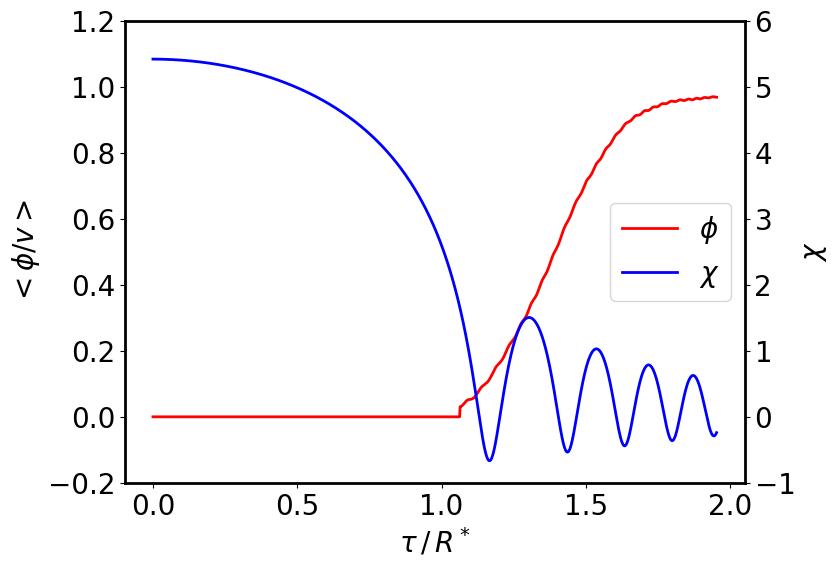}
    \caption{Evolution of the cosmological parameters and field averages during the expansion. The panels illustrate the scale factor $a$ (top-left), the comoving Hubble parameter $\mathcal{H}$ (top-right), the equation-of-state parameter $w$ (bottom-left), and the mean values of the scalar fields $\chi$ and $\phi$ (bottom-right) as functions of time. At the initial time $\tau/R^*=0$, the $\chi$ field begins its evolution from a Gaussian random field with an initial mean value of 5.42, while the $\phi$ field is set to zero. As the system evolves to $\tau/R^*=1.1$, the mean value of the $\chi$ field decreases to 1.69, triggering a phase transition in the $\phi$ field that results in bubble nucleation. Subsequently, the mean value of the $\phi$ field increases, while the $\chi$ field oscillates around $\chi=0$ with a gradually diminishing amplitude. After $\tau/R^*=1.95$, the bubbles in the $\phi$ field have expanded to permeate the entire simulation volume.}
\label{Fig:a,H,omega,field mean}
\end{figure}

We perform simulations on a three-dimensional lattice with resolution $N^3 = 512^3$. The comoving box size $L$ is chosen such that $L\mathcal{H}_0 = 2$, where the initial comoving Hubble parameter $\mathcal{H}_0 = \sqrt{8\pi a_0^2 \rho_{\mathrm{tot}}/3} \simeq 5.5 \times 10^{-6} M_{\mathrm{Pl}}$ corresponds to the GUT scale and is related to the physical Hubble parameter by $\mathcal{H} = aH$. For a given total energy density $\rho_{\mathrm{tot}}$, the initial scale factor $a_0$ is determined self-consistently by this relation. Hereafter, the conformal time $\tau$ has always been subtracted by its initial value $\tau_0$.
For numerical stability, we set the time step to $\mathrm{d}\tau = \mathrm{d}x/5$. The simulation is initialized with the $\chi$ field as a Gaussian random field with mean $\chi = 5.42\,M_{\mathrm{Pl}}$, corresponding to $N = 60$ and $\tau/R^* = 0$. As the $\chi$ field evolves and rolls down to $\chi = 1.69\,M_{\mathrm{Pl}}$ (corresponding to $N = 0.91$), bubbles are nucleated in the $\phi$ field with a number density of $n_b(t_i)/H^3 = 4$ per comoving Hubble volume, following the prescription in Eq.~(\ref{Eq:n_b}). These bubbles subsequently expand and collide until they fill the entire simulation volume. The characteristic bubble separation is given by $R^* = (L^3/N_b)^{1/3}$, where $N_b$ denotes the total number of bubbles.

\begin{figure*}[!htp]
    \centering
    \includegraphics[width=0.24\textwidth]{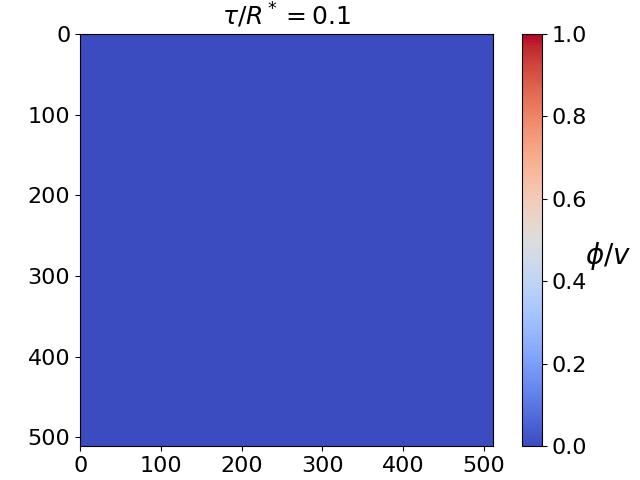}
    \includegraphics[width=0.24\textwidth]{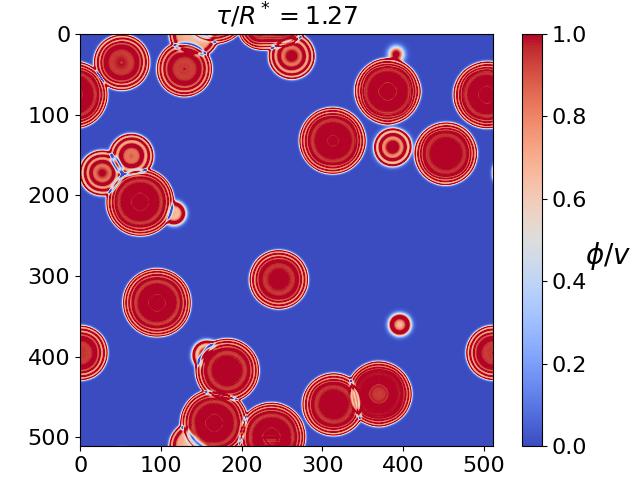}
    \includegraphics[width=0.24\textwidth]{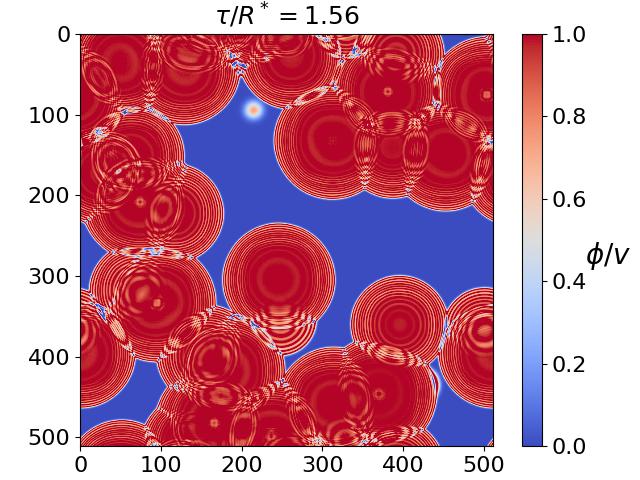}
    \includegraphics[width=0.24\textwidth]{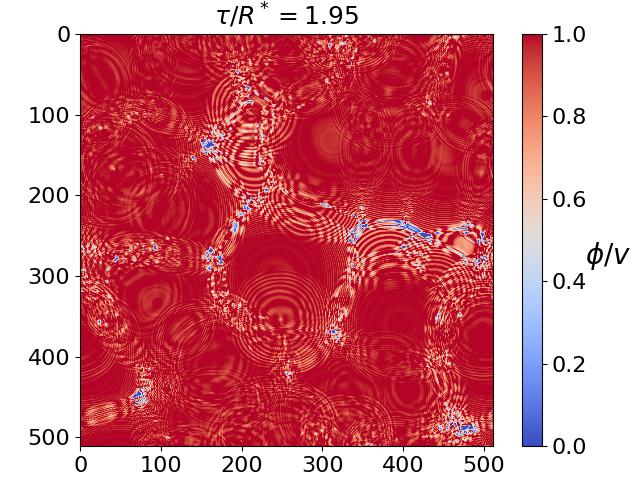}
    \includegraphics[width=0.24\textwidth]{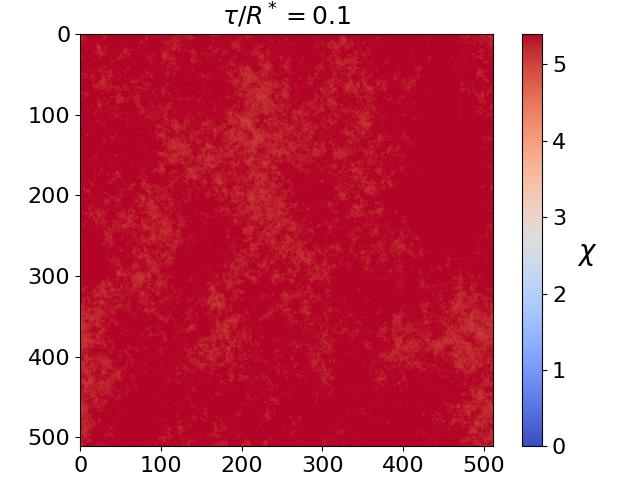}
    \includegraphics[width=0.24\textwidth]{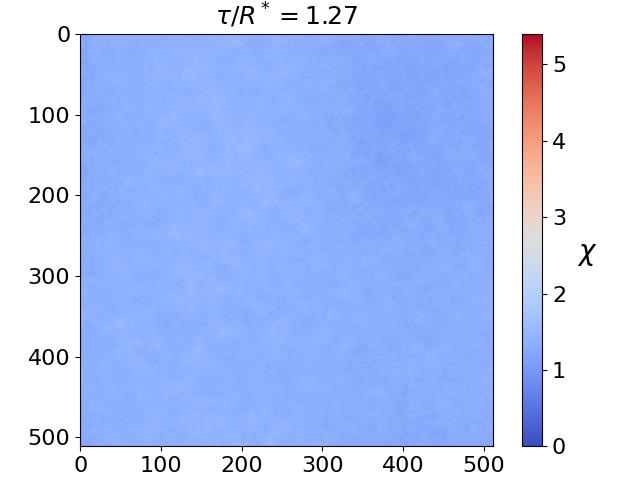}
    \includegraphics[width=0.24\textwidth]{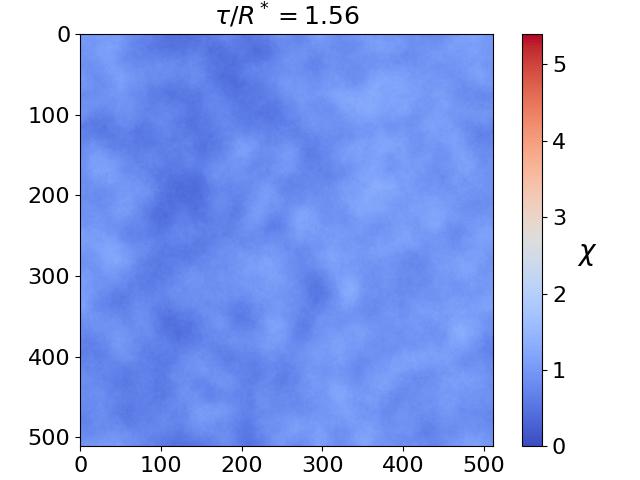}
    \includegraphics[width=0.24\textwidth]{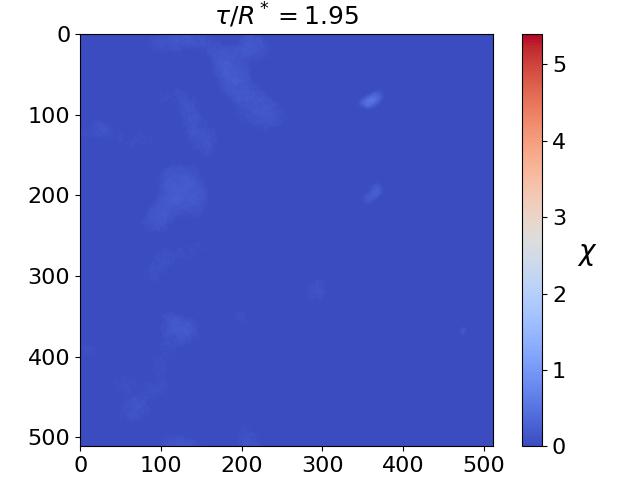}
    \caption{The 2d slices for bubbles dynamic of $\phi$ (top panels) and $\chi$(bottom panels) at different time.}
\label{Fig:2d slices for bubbles}
\end{figure*}

\subsection{Simulation results}
\label{Simulation results}
In this section, we present the numerical simulation results of the model. 
The dynamical process of the FOPT is shown by the evolution of the field averages shown in the bottom-right panel of Fig.~\ref{Fig:a,H,omega,field mean}. Concurrently, upon entering the reheating regime, the $\chi$ field undergoes oscillations around the vacuum minimum at $\chi=0$ before asymptotically vanishing. To further elucidate the macroscopic impact of these field dynamics on the cosmological background, Fig.~\ref{Fig:a,H,omega,field mean} additionally illustrates the evolution of the scale factor $a$, the comoving Hubble parameter $\mathcal{H}$, and the equation of state (EoS) $w$. Our numerical results demonstrate that $\mathcal{H}$ remains nearly constant during the initial stages of the phase transition, subsequently declining after the onset of bubble nucleation at $\tau/R^* \approx 1.1$. In the early epoch, the EoS is maintained at $w = -1$, characteristic of a vacuum-energy-dominated phase. As the nucleation and collision processes unfold, the EoS gradually increases to approximately $-0.6$ and exhibits pronounced oscillatory behavior, driven by the coupling between $\phi$ and $\chi$ alongside the persistent oscillations of the inflaton field.

Fig.~\ref{Fig:2d slices for bubbles} displays representative spatial snapshots of the scalar fields $\phi$ and $\chi$ during the phase transition. Driven by the slow-roll potential, the $\chi$ field initially descends slowly before accelerating. When $\chi$ reaches the critical value of $1.69$, the potential barrier for $\phi$ is sufficiently lowered to facilitate quantum tunneling, triggering the nucleation of vacuum bubbles that subsequently expand, coalesce, and eventually percolate through the entire simulation domain. 

\begin{figure}[!htp]
\centering
    \includegraphics[width=0.48\textwidth]{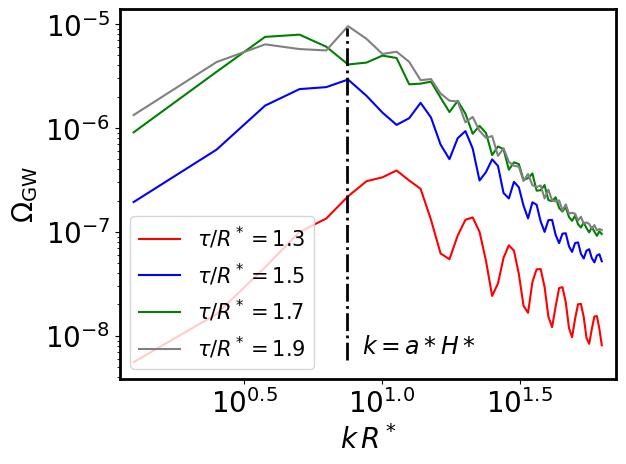}
     \caption{Evolution of the GW energy density spectrum as a function of the comoving wavenumber $k$. The power spectra are shown at four representative moments, capturing the full progression from the onset of the phase transition to the late-stage evolution, where the black vertical line shows the value of k at the phase transition. The GW signal grows with bubble nucleation and collisions, eventually reaching saturation around $\tau/R^* = 1.9$. Notably, a distinct oscillatory structure is observed in the high-frequency regime ($\text{log}_{10}[kR^*] > 1$), which is attributed to the phase accumulation of modes as they exit and re-enter the Hubble horizon.}
\label{Fig:Gws}
\end{figure}

We then turn to the properties of the GW spectrum in this model. Fig.~\ref{Fig:Gws} presents the GW power spectra at four representative moments. As bubbles nucleate and collide within the $\phi$ field, GWs begin to be generated. Here, we define the phase transition time as the moment when the bubble volume fraction reaches 0.7~\cite{Ellis:2018mja,Liu:2021svg}, which is indicated by the vertical line. The signal eventually stabilizes around $\tau/R^* = 1.95$. Notably, in the high-frequency regime where $\text{log}_{10}[kR^*] > 1$, the power spectrum exhibits a distinct oscillatory pattern. This numerical finding is in agreement with the theoretical predictions for inflation-triggered FoPTs~\cite{An:2022cce}. During the early stage, the universe is dominated by vacuum energy, and $\mathcal{H}$ remains relatively stable with little variation. In this background, the GW modes generated by the phase transition are stretched beyond the Hubble horizon due to the rapid expansion and become ``freeze out''. Because modes with different wavelengths exit the horizon at different times, they accumulate distinct phases during the subsequent evolution. Upon re-entering the horizon, the coherent superposition of these modes leads to the characteristic oscillatory structure observed in the power spectrum.
\begin{figure*}[!htp]
\centering
    \includegraphics[width=1.0\textwidth]{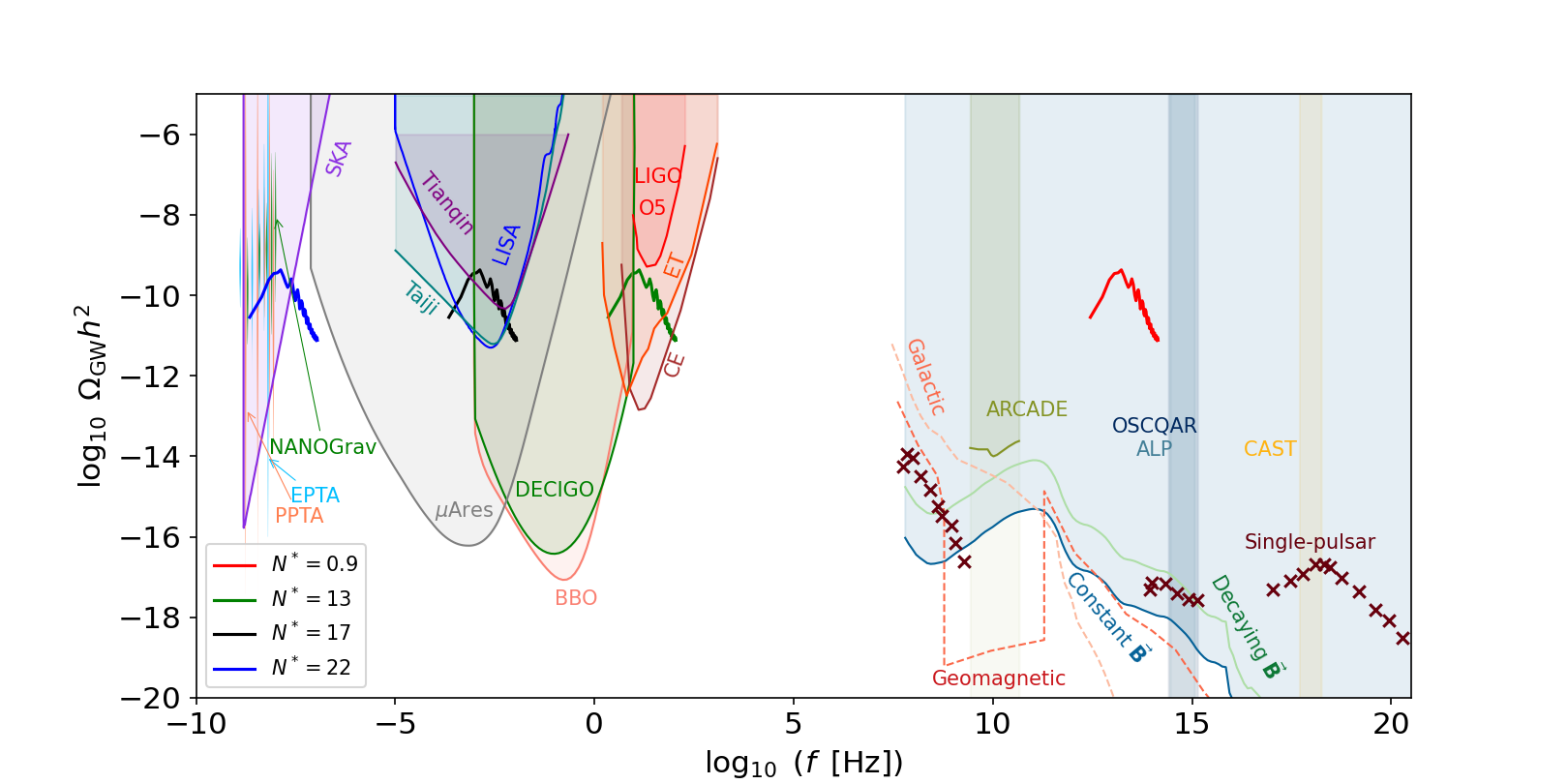}
     \caption{GW spectra from GUT phase transition on different scenarios $N^*=0.9, 13, 17, 22$. The sensitivity curves data comes from~\cite{Hu:2017mde,Ruan:2018tsw,LISA:2017pwj,Baker:2019nia,Crowder:2005nr,Corbin:2005ny,Harry:2006fi,Seto:2001qf,Kawamura:2006up,Yagi:2011wg,Isoyama:2018rjb,EPTA:2015qep,Shannon:2015ect,NANOGRAV:2018hou,Sesana:2019vho,Carilli:2004nx,LIGOScientific:2014pky,Reitze:2019iox,Punturo:2010zz,Cyburt:2015mya,Ejlli:2019bqj,Domcke:2020yzq,Ito:2023nkq,Dandoy:2024oqg}.}
\label{Fig:Gw_red}
\end{figure*}
 
After being redshifted to the present epoch, the peak frequency of the gravitational wave (GW) signal is given by $f_p = 1.6 \times 10^{-5} \frac{\beta}{H_*} \frac{T_*}{100} (\frac{g_*}{100})^{1/6} \text{ Hz}$, with a corresponding spectral amplitude of $\Omega(f_p) = 1.67 \times 10^{-5} (\frac{100}{g_*(T_*)})^{1/3} (\frac{\alpha}{1+\alpha})^2 (\frac{H_*}{\beta})^2$. The resulting redshifted GW spectra are presented in Fig.~\ref{Fig:Gw_red}, alongside the sensitivity curves of various GW detectors. Furthermore, by employing the model from Ref.~\cite{Hu:2025xdt}, When the phase transition is treated as a transient or instantaneous source occurring at the inflationary moment $N^*$, its peak frequency scales as $f_p \propto e^{-N^*}$. we can achieve an earlier phase transition with $N^* = 13, 17, \text{ and } 22$. Consequently, the generated gravitational waves undergo a more pronounced redshift, shifting the signal into lower-frequency bands accessible to current and future detectors.

\section{Conclusions and discussions}
\label{sec:condis}
In this work, we have numerically simulated a specific FoPT at the GUT scale within the Starobinsky inflation framework. While ending inflation through vacuum decay is typically considered infeasible due to insufficient bubble nucleation rates, our approach introduces a potential barrier that evolves exponentially under the dynamic control of the rolling inflaton. This construction effectively suppresses bubble nucleation during the early inflationary era while triggering a massive and rapid burst of nucleations as inflation approaches its end.

The most significant result of this study is the robust confirmation of distinctive oscillatory features in the GW energy density spectrum via full 3D lattice numerical simulations. These high-frequency oscillations successfully reproduce the analytical estimations derived from the ``instantaneous source approximation'', thereby also validating the physical consistency of this mechanism in our specific model. Our simulations demonstrate that the phase transition dynamics are sufficiently rapid and the bubble density sufficiently high to maintain phase coherence during collisions, even when non-linear effects are fully incorporated. These spectral oscillations provide a unique signature for inflation-triggered FoPTs, distinguishing them from standard thermal transitions.

Several aspects merit further improvements and investigations in future works: First, numerical simulations of the whole inflationary era, not just near the end of inflation, are necessary for a rigorous investigation of this two-field dynamics; Second, successive bubble nucleations near the end of inflation, not just at one characteristic moment (the moment that nucleated bubbles collide within one e-folding number), are necessary for a complete survey of bubble collisions, including the collisions between large/rare and small/numerous bubble collisions; Third, particle production might occur by bubble collisions during inflation~\cite{Shakya:2023kjf,Mansour:2023fwj,Inomata:2024rkt}, and even if bubbles cannot collide during inflation for some FoPT model buildings, it is still of great observational interest to study the primordial black hole formation~\cite{Deng:2017uwc,Deng:2020mds,Konoplich:1999qq,Khlopov:2000js,Khlopov:2000js,Dymnikova:2000dy,Khlopov:2024nqp} and isocurvature perturbations~\cite{Buckley:2024nen}.

\begin{acknowledgments}
This work is supported by the National Key Research and Development Program of China (Grants No. 2021YFC2203004, No. 2021YFA0718304, and No. 2020YFC2201501), the National Natural Science Foundation of China (Grants No. 12422502, No. 12547110, No. 12588101, No. 12235019, No. 12547101, and No. 12322505), and the China Manned Space Program (Grant No. CMS-CSST-2025-A01). L.B. also acknowledges the support from the Chongqing Natural Science Foundation (Grant No. CSTB2024NSCQ-JQX0022) and the Chongqing Talents: Exceptional Young Talents Project (No. cstc2024ycjh-bgzxm0020).
\end{acknowledgments}

\bibliography{reference}

\end{document}